# AMAS: optimizing the partition and filtration of adaptive seeds to speed up read mapping

Ngoc Hieu Tran[1,*]
Email: nhtran@ntu.edu.sg

Xin Chen[1]
Email: chenxin@ntu.edu.sg

[1] School of Physical and Mathematical Sciences, Nanyang Technological University, Singapore

* Corresponding author

**ABSTRACT**

**Background:** Identifying all possible mapping locations of next-generation sequencing (NGS) reads is highly essential in several applications such as prediction of genomic variants or protein binding motifs located in repeat regions, isoform expression quantification, metagenomics analysis, etc. However, this task is very time-consuming and majority of mapping tools only focus on one or a few best mapping locations.

**Results**: We propose AMAS, an alignment tool specialized in identifying all possible mapping locations of NGS reads in a reference sequence. AMAS features an effective use of adaptive seeds to speed up read mapping while preserving sensitivity. Specifically, an index is designed to pre-store the locations of adaptive seeds in the reference sequence, efficiently reducing the time for seed matching and partitioning. An accurate filtration of adaptive seeds is further applied to substantially tighten the candidate alignment space. As a result, AMAS runs several times faster than other state-of-the-art read mappers while achieving similar accuracy.

**Conclusions:** AMAS provides a valuable resource to speed up the important yet time-consuming task of identifying all mapping locations of NGS reads. AMAS is implemented in C++ based on the SeqAn library and is freely available at https://sourceforge.net/projects/ngsamas/.

# Background

Recent advances in next-generation sequencing (NGS) technologies have produced massive amounts of short reads data, bringing up promising opportunities in several biomedical research areas such as RNA-seq, ChIP-seq, *de novo* genome sequencing, resequencing, metagenome sequencing, etc [1]. In many applications, the first key step is to map NGS reads to a reference sequence of interest. Thus, dozens of rapid alignment algorithms have been developed to improve the read mapping [2, 3], especially to address the problem of errors tolerance (mismatches, indels) and the increasing read length. To achieve an optimal combination of speed, accuracy, and memory-efficiency, many popular mapping tools such as Bowtie 1, 2 [4, 5], BWA [6-8], etc, apply heuristic approaches to prune the search space, prioritize candidate locations, and finally return one or a few best mapping locations for each read. They can be referred to as best-mappers. Another class of mapping tools, which can be referred to as all-mappers, are specialized in identifying as many as possible, if not all, matches within a reasonable time. This all-mapping task is highly essential in several applications such as prediction of genomic variants or protein binding motifs located in repeat regions, isoform expression quantification, metagenomics analysis [9-12]. Some prominent all-mappers include SOAP 2 [13], SHRiMP 2 [14], mrsFAST [10, 15], mrFAST [9, 16], RazerS 3 [17], GEM [18], Masai [19], Hobbes 1, 2 [20, 21].

Exhaustive search for all mapping locations of NGS reads in a reference sequence is a computationally intensive task. The problem is even more complicated for large genomes with highly repetitive regions, e.g. the human genome, and when sequencing errors as well as genomic variants are taken into account. In order to narrow down the search space, most all-mappers are designed with the seed-and-extend strategy. In particular, a read is first partitioned into a few non-overlapping seeds. If the read can be mapped to the reference sequence with some allowed errors, its seeds must also have exact/approximate matches in the reference sequence, and moreover, the read's mapping locations can be found in the neighbourhood regions of its seeds' locations. Hence, the seeds are quickly mapped to the reference sequence and their identified locations are then used as candidates for further extending the alignment to the rest of the read by using standard dynamic programming algorithms such as Smith-Waterman [22] or Needleman-Wunsch [23]. The mapping sensitivity is guaranteed by the pigeonhole principle. For example, to map a read to the reference sequence with up to $e$ errors, one can partition it into $e+1$ non-overlapping seeds. By the pigeonhole principle, at least one of those $e+1$ seeds must have an exact match in the reference sequence. Thus, all possible mapping locations with up to $e$ errors of the read can be found by extending all exact matches of those seeds.

Most all-mappers partition a read into seeds of equal lengths, where the seed length is specified by users or may also be determined from the number of errors allowed and the pigeonhole principle. In any cases, the seed length is several times shorter than the read length and the seeds are usually mapped exactly (i.e. without errors) to the reference sequence. To further speed up the seed mapping process, most all-mappers (except for RazerS 3) index the reference sequence using some special data structures. In particular, SHRiMP 2, mrsFAST, mrFAST, and Hobbes use hash tables to index the reference sequence. For a fixed length *k* (which is less than or equal to the seed length), the hash tables contain all possible *k*-mers as the keys and each key is associated with a list of locations where the corresponding *k*-mer is observed in the reference sequence. Masai, on the other hand, constructs a suffix tree of the reference sequence to perform the seed matching. Both suffix array and the much more memory-efficient Ferragina-Manzini (FM) index [24] with Burrows-Wheeler Transform (BWT) can be used in Masai to mimic the suffix tree top-down traverse. Hash tables can also be combined to accelerate the searching in the BWT-FM index, as demonstrated in SOAP 2. Among those commonly used data structures, the FM index is well-known for its small memory footprint, which is only ~3-4 GB for the human genome. Interestingly, in the latest version of mrsFAST [15], the authors proposed a new compact structure for hash tables significantly reducing the index size to ~2GB for the human genome, which is even smaller than the FM index.

In addition to indexing the reference sequence, some all-mappers also index the input reads so that multiple reads can be handled simultaneously rather than in a one-by-one fashion. For instance, in the read index of mrsFAST, each *k*-mer is associated with a list of read IDs, locations of the *k*-mer in the reads, etc. In the extension step, for each *k*-mer, the two lists (from the reference index and from the read index) are compared in a divide-and-conquer fashion to achieve cache efficiency. Similarly, in Masai, all seeds of the input reads are organized in a radix tree. Then, by walking through the reference suffix tree along with the read radix tree, all seeds can be mapped to the reference sequence simultaneously.

After the seeds of a read have been mapped to the reference sequence, their identified locations are then used as candidates for further extending the alignment to the rest of the read. This step is usually called the extension or the verification step, and is performed by using standard dynamic programming algorithms such as Smith-Waterman [22] or Needleman-Wunsch [23]. In fact, this step is the bottleneck and consumes the most computations of the whole mapping task. As the seed length is several orders of magnitudes shorter than the reference length and the reference sequence often contains highly repetitive regions, the candidate space generated by the seeds is huge. Moreover, majority of the locations actually are false positives which will not lead to acceptable alignment results yet waste computing resources. Hence, filtration strategies are needed to reduce the number of locations

The first filtration strategy widely used by the all-mappers is based on the pigeonhole principle. In particular, if a read is aligned with up to *e* errors and exact seed matching is employed, only $e+1$ non-overlapping seeds are required to guarantee the full mapping sensitivity. More advanced strategies are further implemented in different all-mappers. For instance, the FastHASH algorithm in the latest version of mrFAST [16] applies two procedures called Cheap *K*-mer Selection and Adjacency Filtering to reduce the number of locations while preserving sensitivity. In the first procedure, the seeds of a read are sorted according to their frequencies in the reference sequence and $e+1$ seeds with the lowest frequencies are selected. The second procedure filters out the obviously false locations based on the idea that the potentially true locations (i.e. those that may return correct alignments) of adjacent seeds must be close to each other in the reference sequence.

In Hobbes [20] the authors proposed a more complicated procedure of seed selection. Instead of partitioning the read into contiguous seeds, Hobbes considers all possible partitions and applies a dynamic programming to select the one that minimizes the sum of frequencies of the seeds. In the latest version, Hobbes 2 [21], the authors further suggested to use extra seeds to improve filtering specificity: instead of $e+1$, $e+2$ (or more) non-overlapping seeds can be employed. Then, by the pigeonhole principle, at least two of those seeds must have exact matches in the reference sequence and hence only those locations reported by at least two seeds will be selected as candidates for the extension step. This technique has also been applied previously in GSNAP [25].

The filtration in Masai [19] is based on the use of long approximate seeds rather than short exact seeds to achieve better specificity. For example, if a read of length 100bp is aligned to the reference sequence with up to $e=5$ errors, by the pigeonhole principle, it can be partitioned into 6 non-overlapping seeds of length 16bp and the seeds are mapped exactly to the reference genome. To increase the seed length, Masai partitions the read into 3 seeds of 33bp and each seed is mapped to the reference genome with up to 1 error. The approximate matching of all seeds is performed simultaneously by applying a multiple backtracking algorithm on the reference index and the read index.

It is worth to note that most all-mappers partition a read into equal-length seeds and prefer to use short $k$-mers as seeds. However, the occurrences of equal-length seeds in a reference sequence may not be uniformly distributed and the seed frequencies may differ by orders of magnitudes (see Supplementary Figure S1 for the frequency distribution of 10-mers in the human genome). Highly repetitive seeds may substantially enlarge the candidate space with false positives and hence waste computing resources. This problem was addressed by introducing seed selection mechanisms in mrFAST, Hobbes, or by using long approximate seeds in Masai. However, the seed frequencies are still not fully controllable. A more efficient strategy of seed partition was proposed in GEM [18] using adaptive seeds. Specifically, GEM allows seeds to have variable lengths while restricting the number of candidate locations generated by each seed to be less than a predefined threshold. Figure 1A shows an example of such efficient partition where the number of candidate locations of the read was reduced by 44.66 times by using adaptive seeds instead of equal-length seeds. In general, adaptive seeds provide better control on the candidate space while consuming similar computing resources as equal-length seeds. Similar ideas have also been used in some best-mappers [7, 8, 26, 27] to avoid unnecessary extension for highly repetitive seeds.

As we have briefly reviewed above, intensive efforts have been put into optimizing the use of equal-length seeds in the all-mapping task, including sequence indexes for fast exact/approximate seed matching and filtration strategies to reduce candidate locations. However, little attention has been paid into adaptive seeds and their advantages have not been fully explored to speed up the read mapping. Hence, in this work we carefully study the strengths and weaknesses of adaptive seeds and attempt to find an optimal strategy for their mapping, partition and filtration. The main contributions of our all-mapping tool with adaptive seeds, AMAS, are summarized below.

Firstly, to speed up the mapping of adaptive seeds, we pre-compute all possible adaptive seeds and their locations in the reference sequence for a given frequency threshold. Then, our index of the reference sequence consists of all possible adaptive seeds as the keys and each key is associated with a list of locations where the corresponding seed is observed. The index is stored in a local file and can be reused for every mapping task. The idea is similar to hash tables, but instead of fix-length *k*-mers, the keys in our index are variable-length substrings with frequencies less than a predefined threshold.

Secondly, in the adaptive seed partition, the number of seeds is not controllable and varies across different reads. For some reads, there may not be sufficient seeds for the pigeonhole principle to guarantee the mapping sensitivity. Hence, we refine the partition of adaptive seeds by adding one more constraint on the seed lengths in order to guarantee the minimum number of seeds in the partition. On the other hand, there may also be more than enough seeds available from the partition. In such cases, we make effective use of the extra seeds to filter out false positives from the candidate space.

Thirdly, due to the shortage of nucleotides assigned to the last seed in the adaptive partition of a read, that seed may be too short and the number of candidate locations it reported may far exceed the desired threshold. Majority of such locations are false positives which increase the candidate space dramatically. However, blindly ignoring all of the last seeds will lead to a considerable loss of mapping sensitivity. Hence, we also pay special attention to an accurate filtration of the last seeds.

Overall, by optimizing the partition and filtration of adaptive seeds, our tool AMAS runs several times faster than other state-of-the-art all-mappers while achieving comparable sensitivity and accuracy. Detailed methods and performance results are presented in the next sections.

AMAS was implemented in C++ based on the source code of Masai and the SeqAn library. Our main contributions include the index, the partition, and the filtration of adaptive seeds. For the extension step with edit distance, we borrowed Masai's implementation [17, 19] of the Myers' bit-vector dynamic programming algorithm [28]. We also borrowed I/O components of Masai for handling the reference, the read sequences, and the alignment results in SAM format. AMAS is freely available at https://sourceforge.net/projects/ngsamas/. A user guide and sample data sets are also provided.

## Methods

To find an effective strategy for the partition and filtration of adaptive seeds, we first performed experiments on a data set of 100k reads of length 100bp, which were simulated from the human genome (UCSC hg19) using Mason [29] with the Illumina model. The reads were mapped to the human genome with up to *e*=5 errors. In the following, we shall describe our strategy and use the simulated data set for illustration.

**Partition of adaptive seeds**

An example of adaptive seeds versus equal-length seeds is illustrated in Figure 1A. To partition a read into adaptive seeds, we scan the read against the reference sequence in the forward (left-to-right) direction, base by base. When the number of matches of the current seed drops below a predefined frequency threshold *F*, its locations are added to the candidate space and a new seed is started. That process is continued until the desired number of seeds is obtained or the read's end is reached. In the latter case, the number of matches of the last seed might be still higher than the threshold *F*. In the example in Figure 1A, the read was mapped to the human genome and *F* was set at 300. The number of candidate locations of the read was reduced by 44.66 times by using 6 adaptive seeds instead of 6 seeds of equal lengths 16bp. Similarly, for the whole simulated data set of 100k reads, the total number of candidate locations was reduced by 6.61 times (Supplementary Table S1).

**Index of adaptive seeds**

To search for the locations of adaptive seeds in the reference sequence, GEM uses the FM index to achieve both speed and low memory footprint [18]. As all-mapping is a very time-consuming task, it may be desirable to further boost up the mapping speed at the expense of a reasonable amount of memory, like using hash tables in Hobbes, mrsFAST, mrFAST, or suffix arrays, enhanced suffix arrays in Masai. Here we pre-compute all possible adaptive seeds and their locations in the reference sequence for a given frequency threshold *F*. Then, our index of the reference sequence consists of all possible adaptive seeds as the keys and each key is associated with a list of locations where the corresponding seed is observed. The idea is similar to hash tables, but instead of fix-length *k*-mers, the keys in our index are variable-length substrings with frequencies less than a predefined threshold *F*.

To represent all possible adaptive seeds in the reference sequence, we implement a lookup table and extend it with a radix tree (Figure 1B). In particular, we first pick a length *k*, organize all *k*-mers in the lookup table and search for their locations in the reference sequence. Then, for each *k*-mer, we further extend the search to its 4 child nodes which correspond to adding the 4 bases "A", "C", "G", and "T". The search and extension are continued until the numbers of matches of the nodes drop below the threshold *F* (shaded leaf nodes in Figure 1B). Thus, all leaf nodes in the radix tree correspond to all possible adaptive seeds and each of them is linked to an array of its locations in the reference sequence. A similar idea has been used in SOAP 2 [13] where a hash table was combined to accelerate the searching in the FM index. Note that the leaf nodes in our index represent adaptive seeds with desired frequencies, not suffixes like in common suffix-array based data structures.

By default, we consider adaptive seeds with a minimum length of *k*=10 bases. In general, using lower values of *F* helps to tighten the candidate space reported by the seeds but also enlarges the index. We choose *F* to be about 10 times less than the average frequency of *k*-mers in the reference sequence. For instance, we set *F*=300 for the human genome (Figure 1B), *F*=10 for the worm genome, and *F*=17 for the fruit-fly genome (Supplementary Table S2). The index only needs to be built once for each genome and it can be stored on a local disk for future mapping tasks.

**Constraint on the maximum seed length**

As we use exact seed matching, by the pigeonhole principle, at least $e+1$ seeds are needed to achieve full sensitivity for mapping a read with up to e errors. The number of adaptive seeds, however, is not controllable and varies across different reads (see Supplementary Figure S2 for the seed count's distribution of the simulated data set). Some reads may not have sufficient seeds to guarantee the mapping sensitivity. Hence, in addition to the frequency constraint, we also impose a constraint on the maximum length of the seeds. The maximum length threshold $l$ allows to control the minimum number of seeds in the partition and hence to improve the mapping sensitivity. For example, for a read of length 100bp, setting $l$=33 guarantees that there will be at least 3 seeds in the partition and all matching locations with up to 2 errors will be found (Supplementary Figure S2). This feature greatly improves the sensitivity of our tool over GEM, as we will show later in the section Results.

**Filtration using the last seeds in the adaptive partition**

We also pay special attention to the last seeds in the adaptive partition. Due to the lack of nucleotides remaining, the last seed in the adaptive partition of a read may be too short and its number of matches may far exceed those of the other seeds as well as the threshold $F$. For instance, in the simulated data set, the number of candidate locations reported by the last seeds contributed up to 64% of the total candidate space. Apparently, majority of those locations were false positives, but removing all of them may lead to a considerable loss of mapping sensitivity. Our tool chooses to filter out only those last seeds whose numbers of matches are higher than the expected frequency of $k$-mers and contribute to more than 95% of the candidate locations of their respective reads (i.e. the last seed reports too many locations compared to other seeds in the same read). Nevertheless, we avoid removing the last seeds from those reads that have too few seeds required by the pigeonhole principle. For the simulated data set, this filtration step only affected a small amount of reads (1.6%, Supplementary Figure S3), but significantly reduced the total candidate space by more than 50% (Supplementary Table S1).

**Filtration using the extra seeds in the adaptive partition**

Last but not least, when there are more than $e+1$ seeds available from the adaptive partition, we make effective use of the extra seeds to filter out false positives. For instance, the simulated reads were mapped with up to $e$=5 errors and hence $e+1$=6 seeds were required. However, the reads could be partitioned up to 9 seeds, where majority of them were partitioned into 8 seeds (Supplementary Figure S2). If a read was partitioned into 7, 8, or 9 seeds, only those locations reported by at least 2, 3, or 4 seeds, respectively, were selected as candidates. This filtration step further reduced the total candidate space by more than 70% (Supplementary Table S1). Our tool stores candidate locations in a binary search tree that allows quickly sorting and identifying those reported by multiple seeds. Using more extra seeds increases the filtering specificity but also requires longer filtration time. Our experiment with the simulated data set showed that using only one extra seed, i.e. total $e+2$=7 seeds, achieved the best overall mapping time (Supplementary Table S1). This is similar to the observation in Hobbes 2 [21], GSNAP [25].

Overall, the combination of two filtration steps using the last seeds and the extra seeds substantially reduced the candidate space of the simulated data set by more than 85% (Supplementary Table S1). Obviously, such effective filtration will save a lot of computing resources in the extension step.

**Seed extension**

To perform seed extension with edit distance (i.e. including both mismatches and indels), we reused Masai's implementation [17, 19] of the Myers' bit-vector dynamic programming algorithm [28].

**Results and discussion**

In this section, we compare the performance of AMAS and the latest state-of-the-art all-mappers GEM [18], Masai [19], Hobbes 2 [21], mrFAST (with FastHASH) [16], and mrsFAST-ultra [15]. We also included two popular best-mappers Bowtie 2 and BWA in the comparison. The mappers were configured to search for all possible mapping locations with up to $e=5$ errors and then output the alignment results in the SAM format. Details of parameters configuration can be found in the Supplementary Table S3. The experiments were performed on a Linux server with 12 Intel Xeon processors (E5-2640, 2.50 GHz), 64 GB of RAM, CentOS 6.3, GCC 4.4.7. All mappers were tested using one single thread and eight threads, except for Masai and mrFAST which do not support multi-threading.

**Rabema benchmark**

Following previous studies of Masai and Hobbes 2 [19, 21], we used the Rabema benchmark [30] to measure the sensitivity of the mappers. First, RazerS 3 [17] was run in its full-sensitive mode to build the gold-standard set of all possible mapping locations with up to 5 errors. Any other mappers that can guarantee full mapping sensitivity can also be used to build the gold-standard set [30]. The gold-standard set was then used by Rabema to assess the sensitivity of each mapper in three categories "All", "All-best", and "Any-best". For the "All" category, the mappers need to find all mapping locations in the gold-standard set. Similarly, for the "All-best" and "Any-best" categories, the mappers need to find all of the best mapping locations and any of the best mapping locations of the reads, respectively. Note that the term "best" here is in terms of the edit distance between the read and the reference sequence. Some mappers such as Bowtie 2 or BWA have their own scoring schemes to evaluate the mapping quality of the alignments. As original locations were available for the simulated reads, we also assessed the recall and the precision of the mappers on the simulated data set. Recall is defined in Rabema as the fraction of the input reads that were correctly mapped and precision is defined as the fraction of the uniquely mapped reads that were correctly mapped. A read is said to be mapped correctly if its original location was reported by the mapper. A read is said to be mapped uniquely if only one location was reported by the mapper.

**Performance on the simulated data set**

Table 1 shows the Rabema benchmark for mapping the simulated data set of 100k reads to the human genome (UCSC hg19). The column "All" shows that AMAS was able to identify 98.11% of all mapping locations and achieved full sensitivity (100%) for those locations with up to 2 errors. Hobbes 2 achieved the highest sensitivity 99.85%, i.e. 1.74% more sensitive than AMAS, and was followed by Masai (99.83%), mrFAST (99.33%), AMAS (98.11%), GEM (97.65%), and mrsFAST (78.52%). AMAS and GEM had lower sensitivity than Masai, Hobbes 2, and mrFAST, especially when searching for mapping locations with 4 or 5 errors. This is because the adaptive seed partition does not guarantee the number of seeds and hence there may not be sufficient seeds required by the pigeonhole principle. The additional constraint on the maximum seed length introduced in AMAS guaranteed the full sensitivity (100%) for all mapping locations with up to 2 errors and also improved the overall sensitivity for locations with 3, 4, 5 errors. Hence, AMAS was 0.46% more sensitive than GEM. mrsFAST had much lower mapping rate (79.35%) and sensitivity (78.52%) than the others because it only allows mismatches and cannot detect indels.

In the "All-best" and "Any-best" categories, AMAS achieved 99.99% and 100% sensitivity, respectively. That was similar to Hobbes 2 and better than the rest. AMAS also achieved the best recall rate 99.09%, while its precision was 99.99%, the second best among the tools.

The best-mappers Bowtie 2, BWA, and GEM-best only searched for one or a few best mapping locations. Hence, they ran faster than the all-mappers while having lower sensitivity. We noticed that GEM-best outperformed Bowtie 2 and BWA in terms of both running time and sensitivity, demonstrating the advantages of adaptive seeds. We tried to configure Bowtie 2 and BWA as all-mappers (Supplementary Table S3) but their running times were too long and the results were not reported here. We also configured Bowtie 2 in its -$k$ mode to report up to -$k$=100 alignments for each read so that it could finish within a reasonable time. Its sensitivity was 96.04%, i.e. 2.07% lower than AMAS.

**Performance on real data sets**

Next, we run the mappers on a real NGS data set SRR063408 from the 1000 Genomes Project (individual HG01495). The data set includes 25.6 million reads of length 100bp which were generated from the Illumina Genome Analyzer II. Table 2 shows the running time, memory footprint, and Rabema benchmark of mapping the first one million reads to the human genome. AMAS took about 30 minutes to finish the job using one single thread and 5 minutes using eight threads. AMAS was able to map 93.75% of the reads and identified 98.14% of all mapping locations, including nearly all of the best mapping locations (99.98%). AMAS only lose behind Masai and Hobbes 2, which achieved highest sensitivity, 99.93% and 99.90% of all mapping locations, 100% of all best mapping locations. In both single-thread and eight-thread tests, AMAS was more than 3 times faster than Hobbes 2. AMAS was also more than 3 times faster than Masai in the single-thread test (Masai does not support multi-threading). While using the same idea of adaptive seeds, AMAS outperformed GEM in terms of mapping rate (0.06% higher), sensitivity (0.25% higher in the "All" category), and running time (2-7 times faster). This clearly shows the benefits of the optimized partition and filtration implemented in AMAS. AMAS was more than 6 times faster than mrFAST in the single-thread test (mrFAST does not support multi-threading). In this experiment, mrFAST produced invalid SAM output with many inconsistent CIGAR strings, which could not be converted to BAM for further analysis with Rabema. mrsFAST had similar running time as AMAS in the single-thread test but was more than 2 times slower than AMAS in the eight-thread test. AMAS also had better mapping rate (1.37% higher) and sensitivity (1.02% higher in the "All" category) than mrsFAST.

To map the full data set SRR063408 of 25.6 million reads to the human genome, AMAS took about 12 hours using one single thread and 2 hours using eight threads (Table 3). Masai required 31 hours and Hobbes 2 required nearly 50 hours, that is, 2.6 times and 4.2 times slower than AMAS when using one single thread. When eight threads were deployed, Hobbes 2 took about 8 hours, 4 times slower than AMAS. mrsFAST was the fastest in the single-thread test (9 hours), but was 2 times slower than AMAS in the eight-thread test (4 hours). mrFAST could not finish the mapping within 3 days, thus we did not record its results. For this data set, GEM was not able to output its alignment results in the SAM format. Thus we were only able to record its mapping time, which was already 1.5-5 times slower than the total running time of AMAS.

We also performed similar experiments on Caenorhabditis elegans (data set SRR065388 mapped to the worm genome UCSC ce10) and Drosophila melanogaster (data set SRR497711 mapped to the fruit-fly genome UCSC dm3). The results are presented in Supplementary Tables S4-S7. AMAS took 9 minutes to map 36 million reads to the worm genome. AMAS outperformed GEM and mrsFAST in terms of running time, mapping rate, and sensitivity. It was nearly 2 times faster than Hobbes 2, 2.5 times faster than Masai and 4.6 times faster than mrFAST. AMAS fell behind Masai and Hobbes 2 by no more than 0.04% of reads and 1.35% of all mapping locations. Similar results were also obtained for the fruit-fly genome.

Most all-mappers use the seed-and-extend strategy in which locations reported by the seeds of a read are used as candidates for extending the alignment to the rest of the read. Many locations reported by different seeds of a read may actually correspond to one single location of the read in the reference sequence. Thus it is important to mark duplicate candidates to reduce the time for seed extension, and more importantly, to avoid duplicate alignments in the output. We measured the percentage of the true alignments (according to Rabema benchmark) found by the mappers among the alignments they reported. Table 4 shows that AMAS and mrsFAST were the most accurate while Masai reported quite a large number of duplicate alignments.

Finally, we tested the mappers on a longer read length and more errors. A data set of one million reads of length 200bp were simulated from the human genome and then were mapped with up to $e=10$ errors (i.e. 5% error rate). The results are shown in Table 5. AMAS was able to map 99.90% of the reads and identified 99.02% of all mapping locations, including nearly all (99.99%) of the best mapping locations. Masai and Hobbes 2 achieved full sensitivity, but were 4-5 times slower than AMAS. GEM did not scale well with this experiment. In the single-thread test its memory footprint increased dramatically and in the eight-thread test it crashed due to memory errors. mrFAST was 3 times slower than AMAS and produced invalid SAM output. mrsFAST had similar running time as AMAS, but much lower mapping rate (62.56%) and sensitivity (61.42% of all mapping locations) because mrsFAST cannot handle indels.

**Index and memory footprint**

As AMAS processes reads in blocks of one million, the memory foot-print is kept stable regardless of the size of the input NGS data (Tables 2 and 3). To map the full data set SRR063408 of 25.6M 100bp reads to the human genome, AMAS requires 19.5 GB. The memory consumption includes 15.9 GB for the index and 3 GB for the reference sequence. The memory footprint is not well optimized yet and shall be the focus of our future development. Hobbes 2 uses hash tables and requires 16.7 GB. Masai uses suffix arrays and requires 25 GB. GEM uses FM index with much less memory footprint 5.7 GB for mapping. However, the memory footprint of GEM increases significantly or even crashes when outputting the alignments to SAM format (using *"gem-to-sam"* tool) or when handling longer reads with more errors (Tables 3, 5). mrsFAST uses hash tables but only needs 6.6 GB. Note that for smaller data sets (with 1M reads), mrsFAST and mrFAST have very low memory footprint, 2 GB, which is even less than the FM index (Table 2).

Our index of the human genome takes about 3 hours to build, and it can be stored on a local disk with 13.8 GB for future mapping tasks (Supplementary Table S2). As our index needs to identify all possible adaptive seeds with desired frequencies and their locations in the reference sequence, it may take more time to build than other indexes. However, this indexing task only needs to be done once for each genome and does not affect the mapping time.

## Conclusions

In this study, we proposed an effective strategy for the partition and filtration of adaptive seeds to speed up the exhaustive search for all possible mapping locations of NGS reads in a reference sequence. Our tool, AMAS, features a new index designed specifically to accelerate the alignment of adaptive seeds to the reference sequence. Frequency and length constraints on adaptive seeds allow better control on the candidate alignment space while preserving enough seeds to guarantee the mapping sensitivity. Finally, an accurate filtration strategy based on the last seeds and the extra seeds is further applied to substantially tighten the candidate alignment space, efficiently reducing the time for seed extension. As a result, AMAS runs several times faster than other competitors while achieving comparable sensitivity and accuracy.

Multi-mapping reads are highly essential in several applications such as prediction of genomic variants or protein binding motifs located in repeat regions, isoform expression quantification, metagenomics analysis, etc. However, searching for them is a very time-consuming task. Current state-of-the-art all-mappers such as GEM, Masai, Hobbes 2, mrsFAST, mrFAST, and best-mappers such as Bowtie 2, BWA, etc, may take up to 1-2 days to find all possible mapping locations of reads from a single NGS run for the human genome. Our tool offers a more efficient way to significantly speed up that process. We have shown that AMAS run 2-7 times faster and was more sensitive than GEM.

AMAS was also up to 4 times faster than Masai, Hobbes 2, and mrFAST. Although AMAS did not achieve nearly full sensitivity like Masai, Hobbes 2, and mrFAST, it only missed out less than 1.76% of all mapping locations and 0.02% of all best mapping locations, while guaranteeing full sensitivity for matches with small numbers of errors. Most of missing locations by AMAS were those with high numbers of errors and may not be of interest in practice.

AMAS scales well in multi-threading environment and keeps the memory footprint stable regardless of the size of input data. AMAS also supports pair-end read mapping, best-mapping mode, and $-k$ mapping mode (reporting up to $-k$ alignments for each read). Hence, we believe that AMAS will become a useful resource for read mapping in NGS data analysis.

## Competing interests

The authors declare that they have no competing interests.

## Authors' contributions

XC and NHT designed the project. NHT developed the software and performed the analysis. NHT and XC wrote the manuscript. Both authors read and approved the manuscript.

## Acknowledgements

We thank the SeqAn team, especially the authors of Masai package for making their source codes available. Our tool uses the SeqAn library and some well-developed components of Masai.

This work was supported by the Singapore National Medical Research Council (CBRG11nov091) and the Singapore Ministry of Education Academic Research Fund (MOE2012-T2-1-055)

**Figure Legends**

**Figure 1. Partition of adaptive seeds. (A)** Partition of the read "AGGTCAAGAAATTGAGACCATCCCGGCCAACTTGGTGAAACCCCGTCTCGTCTCTACTAAAAGTAAAAAAATTAGCTGGGCGTGGTGGCGGGTGCATAT" against the human genome using 6 equal-length16bp seeds (upper) and 6 adaptive seeds with frequencies less than $F$=300 (lower). Using adaptive seeds significantly reduced the total number of candidate locations by ~44.66 times. **(B)** A partial sub-tree of our index with $k$=10, $F$=300 for the human genome. Nodes were extended until their numbers of matches dropped below 300 (shaded leaf nodes). The path corresponding to the first adaptive seed of the read in (A) with length 13bp and frequency 157 is illustrated with bold lines.

**Tables**

**Table 1. Rabema benchmark for mapping the simulated data set of 100k reads of length 100bp to the human genome with up to *e*=5 errors**. The first seven tools are all-mappers and the last three are best-mappers. The column "All" shows the percentage of all mapping locations that the mappers were able to find. Numbers in bold show the overall percentage, numbers in the smaller font size show detailed percentages for mapping locations with 0, 1, 2 (upper) and 3, 4, 5 (lower) errors. Similarly, the two columns "All-best" and "Any-best" respectively show the sensitivity of the mappers for identifying all of the best mapping locations and any of the best mapping locations of the reads. The last two columns, "Rec" and "Prec", show the recall and the precision of the mappers, respectively (see the main text for the definitions of recall and precision).

| Mappers | Reads mapped (%) | All (%) | | | | All-best (%) | | | | Any-best (%) | | | | Rec (%) | Prec (%) |
|---|---|---|---|---|---|---|---|---|---|---|---|---|---|---|---|
| AMAS | 99.11 | **98.11** | 100 | 100 | 100 | **99.99** | 100 | 100 | 100 | **100** | 100 | 100 | 100 | 99.09 | 99.99 |
| | | | 98.71 | 90.76 | 72.53 | | 99.94 | 99.77 | 99.94 | | 99.97 | 99.89 | 100 | | |
| Bowtie2 (k=100) | 99.06 | **96.04** | 99.97 | 99.82 | 98.73 | **99.59** | 99.71 | 99.69 | 99.34 | **99.88** | 100 | 99.92 | 99.70 | 98.40 | 99.76 |
| | | | 93.02 | 76.36 | 52.47 | | 99.20 | 98.92 | 97.26 | | 99.65 | 99.27 | 97.50 | | |
| GEM (all) | 99.06 | **97.65** | 100 | 99.99 | 99.82 | **99.85** | 100 | 99.87 | 99.78 | **99.92** | 100 | 99.98 | 99.87 | 98.72 | 99.73 |
| | | | 97.27 | 88.26 | 67.40 | | 99.64 | 99.24 | 96.52 | | 99.82 | 99.44 | 96.86 | | |
| Masai | 99.06 | **99.83** | 100 | 100 | 100 | **99.95** | 100 | 100 | 100 | **99.95** | 100 | 100 | 100 | 99.05 | 100 |
| | | | 99.74 | 99.38 | 97.46 | | 99.70 | 99.05 | 98.35 | | 99.70 | 99.05 | 98.47 | | |
| Hobbes2 | 99.11 | **99.85** | 100 | 100 | 100 | **100** | 100 | 100 | 100 | **100** | 100 | 100 | 100 | 98.99 | 99.89 |
| | | | 100 | 99.94 | 97.28 | | 100 | 99.97 | 99.92 | | 100 | 100 | 99.92 | | |
| mrFAST | 98.55 | **99.33** | 100 | 100 | 100 | **99.44** | 100 | 100 | 100 | **99.44** | 100 | 100 | 100 | 98.00 | 99.43 |
| | | | 100 | 99.99 | 87.95 | | 100 | 100 | 55.16 | | 100 | 100 | 55.31 | | |
| mrsFAST | 79.35 | **78.52** | 100 | 74.50 | 53.65 | **79.72** | 100 | 74.09 | 52.27 | **79.76** | 100 | 74.16 | 52.33 | 78.14 | 98.75 |
| | | | 39.40 | 37.93 | 48.97 | | 31.88 | 11.55 | 02.46 | | 31.93 | 11.60 | 02.50 | | |
| Bowtie2 (best) | 98.58 | **90.12** | 97.91 | 96.59 | 92.00 | **96.01** | 96.58 | 96.27 | 94.61 | **99.27** | 100 | 99.47 | 97.53 | 94.88 | 96.25 |
| | | | 79.43 | 51.96 | 20.70 | | 94.15 | 93.14 | 92.20 | | 97.25 | 96.13 | 95.49 | | |
| GEM (best) | 99.06 | **94.19** | 100 | 99.29 | 97.25 | **99.85** | 100 | 99.86 | 99.76 | **99.91** | 100 | 99.96 | 99.87 | 98.61 | 99.70 |
| | | | 90.26 | 67.59 | 32.80 | | 99.54 | 99.21 | 96.50 | | 99.77 | 99.50 | 96.94 | | |
| BWA | 98.04 | **89.43** | 97.91 | 96.68 | 92.71 | **95.53** | 96.58 | 96.54 | 96.21 | **98.76** | 100.00 | 99.68 | 99.20 | 93.92 | 95.79 |
| | | | 75.48 | 42.92 | 14.71 | | 90.13 | 77.24 | 65.54 | | 93.00 | 80.16 | 68.20 | | |

**Table 2. Rabema benchmark, running time, and memory footprint for mapping the first one million reads of the real Illumina data set SRR063408 to the human genome with up to *e*=5 errors.** The read mapping was performed using 1 thread and 8 threads. Masai and mrFAST do not support multi-threading. The column "All" shows the percentage of all mapping locations that the mappers were able to find. Similarly, the two columns "All-best" and "Any-best" respectively show the sensitivity of the mappers for identifying all of the best mapping locations and any of the best mapping locations of the reads. In this experiment, mrFAST produced invalid SAM output with many inconsistent CIGAR strings, which could not be converted to BAM for further analysis. NA: not available.

| Mappers | 1-thread Time (mm:ss) | 8-thread Time (mm:ss) | Memory (GB) | Reads mapped (%) | All (%) | All-best (%) | Any-best (%) |
|---|---|---|---|---|---|---|---|
| AMAS | 29:43 | 04:59 | 19.4 | 93.75 | 98.14 | 99.98 | 99.99 |
| Bowtie2 (k=100) | 55:30 | 07:31 | 3.6 | 93.70 | 95.89 | 99.60 | 99.91 |
| GEM (all) | 71:41 | 36:45 | 5.3 | 93.69 | 97.89 | 99.91 | 99.92 |
| Masai | 91:12 | NA | 16.7 | 93.75 | 99.93 | 100 | 100 |
| Hobbes2 | 110:24 | 15:16 | 14.9 | 93.76 | 99.90 | 100 | 100 |
| mrFAST | 189:47 | NA | 2.0 | NA | NA | NA | NA |
| mrsFAST | 29:41 | 11:28 | 2.1 | 92.38 | 97.12 | 98.48 | 98.50 |

**Table 3. Details of mapping the full data set SRR063408 of 25.6 million reads to the human genome with up to *e*=5 errors.** The read mapping was performed using 1 thread and 8 threads. Masai and mrFAST do not support multi-threading. For this full data set, GEM was not able to output the alignment results in the SAM format. Thus we only recorded its running time and memory consumption for the mapping step, but not the writing step. mrFAST could not finish the mapping within 3 days, thus we did not record its results. NA: not available.

| Mappers | AMAS | Bowtie2 (k=100) | GEM* (all) | Masai | Hobbes2 | mrFAST | mrsFAST |
|---|---|---|---|---|---|---|---|
| 1-thread Time (hh:mm) | 12:09 | 24:23 | 18:07 | 31:02 | 49:56 | NA | 08:48 |
| 8-thread Time (hh:mm) | 01:59 | 03:19 | 10:41 | NA | 08:13 | NA | 04:06 |
| Memory (GB) | 19.5 | 3.8 | 5.7 | 25.0 | 16.7 | NA | 6.6 |

**Table 4. Percentage of the true alignments found by the mappers among the alignments they reported.** For the real data set of one millions reads, mrFAST produced invalid SAM output with many inconsistent CIGAR strings, which could not be converted to BAM for further analysis. NA: not available.

| Mappers | AMAS | GEM (all) | Masai | Hobbes 2 | mrFAST | mrsFAST |
|---|---|---|---|---|---|---|
| *Simulated data set of 100k reads* | | | | | | |
| Number of true alignments found | 8,886,145 | 8,624,695 | 13,813,089 | 13,735,743 | 13,806,005 | 10,785,780 |
| Number of alignments reported | 8,886,270 | 8,628,751 | 21,954,696 | 13,762,313 | 13,918,070 | 10,785,975 |
| Percentage | 100 | 99.95 | 62.92 | 99.81 | 99.20 | 100 |
| *Real data set of 1M reads* | | | | | | |
| Number of true alignments found | 99,339,852 | 105,459,737 | 148,827,680 | 148,526,618 | NA | 134,412,384 |
| Number of alignments reported | 99,447,106 | 105,602,112 | 243,162,835 | 148,743,923 | 149,368,650 | 134,506,255 |
| Percentage | 99.89 | 99.87 | 61.21 | 99.85 | NA | 99.93 |

**Table 5. Rabema benchmark, running time, and memory footprint for mapping the simulated data set of one million reads of length 200bp to the human genome with up to *e*=10 errors.** The read mapping was performed using 1 thread and 8 threads. Masai and mrFAST do not support multi-threading. The multi-threading mode of GEM also crashed for this data set due to memory problem. The column "All" shows the percentage of all mapping locations that the mappers were able to find. Similarly, the two columns "All-best" and "Any-best" respectively show the sensitivity of the mappers for identifying all of the best mapping locations and any of the best mapping locations of the reads. In this experiment, mrFAST produced invalid SAM output with many inconsistent CIGAR strings, which could not be converted to BAM for further analysis. NA: not available.

| Mappers | 1-thread Time (mm:ss) | 8-thread Time (mm:ss) | Memory (GB) | Reads mapped (%) | All (%) | All-best (%) | Any-best (%) | Recall (%) | Precision (%) |
|---|---|---|---|---|---|---|---|---|---|
| AMAS | 38:58 | 05:55 | 19.6 | 99.90 | 99.02 | 99.99 | 100 | 99.89 | 100 |
| Bowtie2 (k=100) | 185:30 | 24:41 | 3.6 | 99.90 | 98.20 | 99.76 | 99.94 | 99.64 | 99.90 |
| GEM (all) | 332:47 | NA | 19.7 | 99.90 | 98.71 | 99.95 | 99.99 | 99.42 | 99.55 |
| Masai | 217:06 | NA | 17.0 | 99.90 | 100 | 100 | 100 | 99.90 | 100 |
| Hobbes2 | 180:35 | 24:21 | 14.2 | 99.90 | 99.91 | 100 | 100 | 99.81 | 99.91 |
| mrFAST | 122:26 | NA | 2.3 | NA | NA | NA | NA | NA | NA |
| mrsFAST | 33:53 | 07:08 | 2.9 | 62.56 | 61.42 | 62.28 | 62.31 | 60.92 | 97.71 |

**Additional files**

Additional file 1 – Supplementary figures

Additional file 2 – Supplementary tables

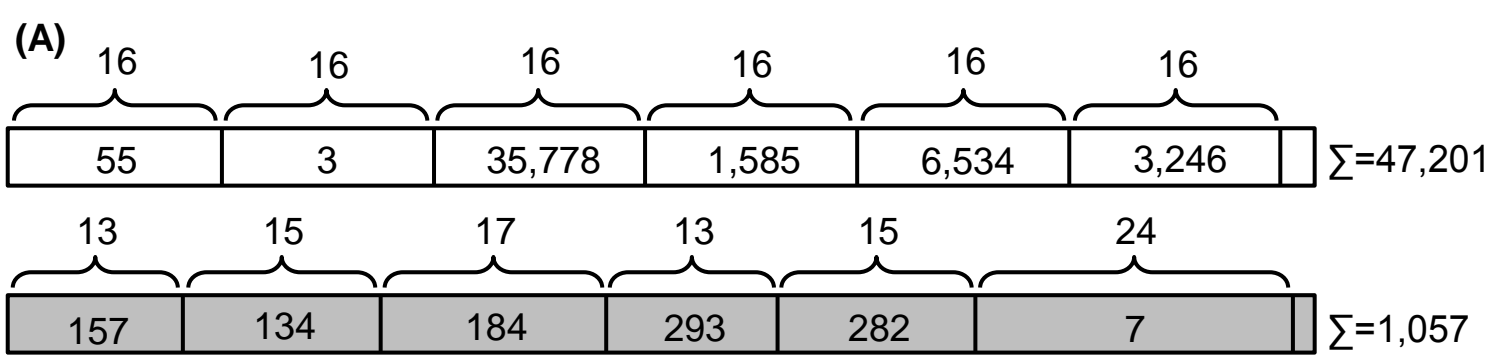
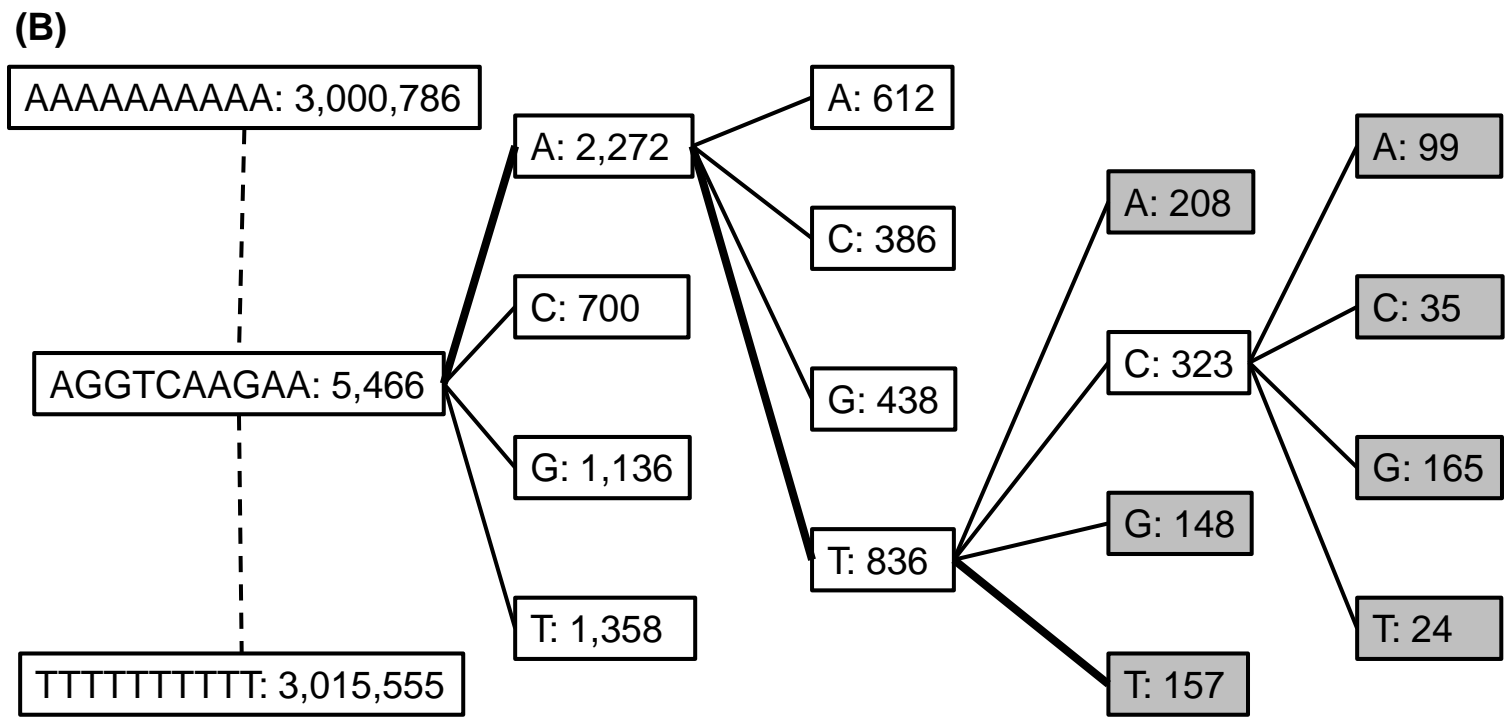